\documentclass[11pt]{article}
\usepackage{latexsym}  
\usepackage{amssymb}
\usepackage{graphicx}
\usepackage{amsmath}

\begin{document}

\begin{center}

{\Large\bf Sumino's Cancellation Mechanism} \\[2mm]

{\Large\bf in an Anomaly-Free Model } 

\vspace{4mm}

{\bf Yoshio Koide}

{\it Department of Physics, Osaka University, 
Toyonaka, Osaka 560-0043, Japan} \\
{\it E-mail address: koide@kuno-g.phys.sci.osaka-u.ac.jp}

\date{\today}
\end{center}

\vspace{3mm}

\begin{abstract}
 An interesting family gauge boson (FGB) model (Model A) has 
been proposed by Sumino. The model can give FGBs with a 
considerably low energy scale in spite of the sever 
constraints form the observed 
$K^0$-$\bar{K}^0$ mixing and so on.  
An essential idea in Model A is in the so-called 
Sumino cancellation mechanism between QED and FGB diagrams. 
However, Model A is not anomaly free and, besides, it causes 
effective interactions with $\Delta N_{\rm family}=2$. 
In order to avoid these problems, a revised Sumino model 
with an inverted mass hierarchy (Model B) has proposed, 
but, in this time, it cannot satisfy the Sumino cancellation 
mechanism exactly.  In this paper, we propose a revised version 
of Model B, where the model still keeps anomaly free, but 
it can exactly satisfy the Sumino mechanism.  
An effect of the revised model will be confirmed by observations 
$K^+ \rightarrow \pi^+ e^- \mu^+$ and $\mu^- N\rightarrow e^- N$.
\end{abstract}

PCAC numbers:  
  11.30.Hv, 
  12.60.-i, 

\vspace{3mm}

\noindent{\bf 1  \  Introduction}

The most challenging subject in particle physics is 
to understand the origin of flavors. 
There is an attractive idea that the flavor physics is 
understood from the point of view of a family symmetry. 
Then, it is natural to consider that the symmetry is gauged,
because if we consider a global symmetry, we will have 
unwelcome massless Nambu-Gloldstone scalars. 
So, we may expect existence of family gauge bosons (FGBs).
However, in the conventional FGB models, it is considered that 
a scale of the family symmetry breaking is extremely high, 
because of constraints from the observed $P^0$-$\bar{P}^0$ 
mixings ($P= K, D, B, B_s$). 
Therefore, the observation of FGBs was not a realistic topic.  

Against such a conventional view,   
in 2009, Sumino \cite{Sumino_PLB09} gave a realistic role 
to FGBs. 
Thereby, Sumino has speculated that a scale of the FGBs 
is of an order of $10^3$ TeV. 

Prior to a review of the Sumino FGB model, let us give a 
brief review of a charged lepton mass relation~\cite{K-mass}, 
$$ 
K \equiv \frac{m_e + m_\mu +m_\tau}{\left(\sqrt{m_e}+ \sqrt{m_\tau} 
+\sqrt{m_\tau}\right)^2} = \frac{2}{3} . 
\eqno(1.1)
$$
The relation is satisfied by the pole masses [i.e.\ $K^{pole}=(2/3) 
\times(0.999989 \pm 0.000014)$], but not so well satisfied by the
running masses [i.e.\ $K(\mu)=(2/3) \times (1.00189 \pm 0.00002)$ at
$\mu=m_Z$].
The formula (1.1) was first derived based on a U(3) family symmetry 
model by the author in 1982. 
However, it should be considered that the formula (1.1) as a result 
from a U(3) model should be distinguished from the excellent 
agreement $K^{pole}= 2/3$.  
The agreement $K^{pole}= 2/3$ may be taken as a accidental 
coincidence irrespective of $K^{model} =2/3$.  

Nevertheless, against our common sense, in 2009, 
Sumino \cite{Sumino_PLB09} has paid attention 
to a fact of the ``coincidence". 
Sumino has tried to understand why the formula (1.1) is 
well satisfied with the pole masses not with the running 
masses. 
He has introduced FGBs, and thereby,
he has given an explanation. 
There, a cancellation mechanism between QED correction 
and  FGB correction terms has been proposed. 
(Hereafter, we will refer his model as Model A.) 
The mechanism will be reviewed in the next section.

However, his model is not anomaly free, and besides, 
the model causes an unwelcome interactions with 
$\Delta N_{fam}=2$ ($N_{fam}$ is family number).
Therefore, a revised version of Model A has been 
proposed \cite{K-Y_PLB12}. 
(Hereafter, we refer it as Model B). 
Although Model B is anomaly free and it has 
no $\Delta N_{fam}=2$ interaction, regrettably, the 
Sumino cancellation in Model B is only approximately 
realized. 

The purpose of the present paper is to propose 
an improved model of Model B. 
(We will refer it as Model C.)
(The purpose is not to derive the relation (1.1) itself, 
nor to propose a revised version of Model B.
For the derivation of (1.1), for example, see 
a scalar potential model by the author \cite{YK_MPLA90}, 
 a $\Sigma(81)$ model by Ma \cite{Ma_PLB07}, 
a U(9) model by Sumino \cite{Sumino_JHEP09}, and so on.)

Thereby, we will re-estimate FGB masses.  
As a result, we will obtain slightly light 
mass values of FGBs compared with Model B. 
Sine family-number violating rare decay width 
due to FGB exchange is proportional to the factor 
$(g_F/M_{ij})^4$, the predicted value of 
the rare decay will changed considerably. 
So, some of modes will become within our reach 
soon.


\vspace{3mm}

\noindent{\bf 2  \  Sumino canellation mechanism} 

In order to give well-understanding of our model  
(Model C) in this section, let us give a brief review 
of the Sumino cancellation mechanism.  

Note that the formula (1.1) is invariant under to a 
transformation 
$$
m_{ei}(\mu) \rightarrow m_{ei} (\mu) \left[
1 + \varepsilon_0(\mu) + \varepsilon_i(\mu) \right] , 
\eqno(2.1)
$$ 
with
$\varepsilon_i =0$, where $\varepsilon_0$ and 
$\varepsilon_i$ are family-number independent and 
dependent factors, respectively.  
The QED correction for the running mass $m_{ei}(\mu)$ is 
given as \cite{Arason92}:
$$
m_{e_i}(\mu) = m_{e_i}^{pole} \left[ 1-\frac{\alpha_{em}(\mu)}{\pi} 
\left( 1 +\frac{3}{4} \log \frac{\mu^2}{m_{e_i}^2(\mu)} \right) \right] .
\eqno(2.2)
$$
Therefore, if the family-number dependent factor 
$\log m_{e_i}^2$ in Eq.(2.2) is absent, then the running 
masses $m_{e_i}(\mu)$  
will also satisfy the formula (1.1). 
Noticing this fact,  
Sumino has proposed a U(3) family
gauge model \cite{Sumino_PLB09}, where a factor $\log m_{e_i}^2$
in the QED correction for the charged lepton mass $m_{e_i}$ 
($i=1,2,3$) is canceled by a factor $\log M_{ii}^2$ in 
a corresponding diagram due to the FGBs.
Here,  Sumino has introduced a scalar $\Phi$ which is 
$({\bf 3}, {\bf 3})$ of U(3)$\times$O(3),
whose symmetry breaking scales are $\Lambda$ and 
$\Lambda'$, respectively ($\Lambda \ll \Lambda'$). 
The vacuum expectation value (VEV) $\langle \Phi \rangle$ is 
given by 
$$
\langle \Phi_i^{\ \alpha} \rangle = [ {\rm diag} (v_{e1} , 
v_{e2} , v_{e3} )]_i^{\ \alpha} , 
\eqno(2.3)
$$
where $i$ and $\alpha$ are indexes of U(3) and O(3), 
respectively. 
(Hereafter, since we consider $\Phi$ of  
$({\bf 3}, {\bf 3}^*)$ of U(3)$\times$U(3)$'$ in 
an extended Sumino model later, 
we have denoted the components of  $\Phi$ as $\Phi_i^{\ \alpha}$,
not  $\Phi_{i\alpha}$.)
Then, the charged lepton mass matrix $M_e$ in the Sumino model 
is given by
$$
(M_e)_i^{\ j} = k_e \langle \Phi_i^{\ \alpha} \rangle
\langle \bar{\Phi}_\alpha^{\ j} \rangle = k_e \delta_i^j v_{ei}^2,  
\ \ \ \ {i.e.} \ \ 
m_{ei} = k_e v_{ei}^2 .
\eqno(2.4)
$$
On the other hand, the scalar $\Phi$ also generates masses of 
FGBs $A_i^{\ j}$ as
$$
M^2_{ij} = \frac{1}{2} g_F^2 \left\{
\left(\langle \Phi \rangle_i^{\ \alpha} 
\langle {\Phi}^\dagger\rangle _\alpha^{\  i} + 
\langle \bar{\Phi}^\dagger \rangle_i^{\ \alpha} 
\langle \bar{\Phi} \rangle _\alpha^{\  i} \right) +
(i\rightarrow j) \right\} 
 \propto v_{ei}^2 +v_{ej}^2 \propto m_{ei} + m_{ej},  
 \eqno(2.5)
 $$
in the limit of $\Lambda' \gg \Lambda$.   
Sumino's idea is as follows:  
The factor $\log m_{ei}$ in (2.2) is canceled by a factor 
$\log M_{ii}$ in  the radiative FGB diagram  
by taking a suitable relation between the electromagnetic gauge 
coupling constant $e$ and the family gauge coupling constant 
$g_F$ and by noticing 
$\log M_{ii}^2 = \log (k_e m_{ei}) =\log m_{ei}  + \log k_e$: 
$$
m_{e_i}(\mu) = m_{e_i}^{pole} \left[ c_{QED}(\mu) \log m_{e_i}(\mu)
+ c_{FGB}(\mu) \log m_{e_i}(\mu)+ C(\mu)   \right] .
\eqno(2.6)
$$
Note that the last term $C(\mu)$ in (2.6) is a flavor-blind 
term and not zero, 
so that the running mass $m_{ei}(\mu)$ still has the 
energy scale dependence even if the cancellation is realized. 
Of course, the contributions to $m_{ei}(\mu)$ are not only 
QED and FGB.  
Although most contributions are flavor blind, so that those 
are absorbed into the term $C(\mu)$ in (2.6), 
some contributions which are not flavor blind still remain 
as small contributions. 
Since the purpose of the Sumino mechanism is not to give 
the exact cancellation among $\log m_{ei}$ terms for 
whole energy scale, but to give the cancellation within 
the present experimental error, 
Sumino has concluded that an applicable energy scale
of the cancellation mechanism (2.6) is at most $10^3$ TeV 
by estimating a deviation from the relation (1.1) 
due to such the small contributions  \cite{Sumino_PLB09}. 
(However, the author guesses that the limit $10^3$ TeV 
should be not taken rigidly and that an order of 
$10^4$ TeV is also allowed.)
Sumino has also told that if the measurement of tau lepton mass 
will be improved by one order, then a deviation from 
the charged lepton mass relation (1.1) will be observed 
\cite{Sumino_private}. 

Note that not only the Sumino model gives a possible 
explanation for well-satisfied relation $K^{pole}=2/3$,  
but also the model gives 
a FGB model with a considerably low energy scale.  
In the Sumino model, the FGB mass matrix is diagonal in the 
diagonal basis of the charged lepton mass matrix, so that 
family-number violation does not occur in the lepton sector. 
Family-number violation occurs only via the quark mixing.   
In the limit of no quark mixing, the family-number violation 
in the quark sector is also forbidden.
Therefore, contribution of FGBs to the $P^0$-$\bar{P}^0$ mixing
can be considerably reduced compared with the conventional FGB 
models. (See Ref.\cite{YK_PLB14}.) 
 
However, in the Sumino model, in order to get 
a minus sign for the purpose of the cancellation, 
leptons $(\nu_i,e^-_i)_L$ and $(\nu_i, e^-_i)_{R}$ 
are assigned to ${\bf 3}$ and ${\bf 3}^*$ of U(3) family 
symmetry, respectively.
Therefore, the Sumino model is not anomaly free.
(In order to avoid this anomaly problem, we may add some 
heavy leptons.   However, then, the model will become 
somewhat complicated.)
Besides, the model inevitably causes effective interaction with 
$\Delta N_F =2$ ($N_F$ is a family number). 

In order to evade these problems, Yamashita and the author 
\cite{K-Y_PLB12} have proposed an extended model, Model B,  
with anomaly free. 
Here, the fermion $(\nu_i,e^-_i)_L$ and 
$(\nu_i, e^-_i)_{R}$ are assigned to ${\bf 3}$ and ${\bf 3}$ 
of U(3) family symmetry, respectively and the FGB masses 
have inverted mass hierarchy  
$$
M_{ij}^2 =K_e(m_{ei}^{-1} + m_{ej}^{-1}), 
\eqno(2.7)
$$
in order to obtain a minus sign for the Sumino cancellation 
mechanism,
i.e. $\log M_{ii}^2 = -\log m_{ei} + \log (2 K_e))$. 
However, in order to give the inverted masses  hierarchy 
(2.7), we must introduce additional scalar $\Psi$ which 
belongs to $({\bf 3}, {\bf 3}^*)$ of U(3)$\times$U(3)$'$ 
and which has a VEV 
$$
\langle \Psi_i^{\ \alpha} \rangle = [{\rm diag}
 (v_{F1}, v_{F2}, v_{F3}) ]_i^{\ \alpha} ,
 \eqno(2.8)
 $$
with a VEV relation
$$
\langle \Psi_i^{\ \alpha}\rangle 
\langle \bar{\Psi}_\alpha^{\ k}\rangle 
\langle \Phi_k^{\ \beta}\rangle 
\langle \bar{\Phi}_\beta^{\ j}\rangle 
= k_\Psi k_\Phi \delta_i^j , \ \ \ \ {\rm i.e.} \ \ \ 
(v_{Fi})^2 =  k_\Psi k_\Phi (v_{ei})^{-2} .
\eqno(2.9)
$$
For convenience, let us denote a superpotential 
which gives the VEV relation (2.9) as follows:
$$
W_{\Phi\Psi}^{eff} = \frac{\lambda_1}{\Lambda^2} \left[ 
(\Psi\bar{\Psi} \Phi \bar{\Phi})_i^{\ j} (\Theta)_j^{\ i} \right] 
+ \frac{\lambda_2}{\Lambda^2} \left[
(\Psi\bar{\Psi} \Phi \bar{\Phi})_i^{\ i} \right] 
\left[ (\Theta)_j^{\ j} \right], 
\eqno(2.10)
$$
where we have assumed that the flavon $\Theta$ always takes 
$\langle \Theta \rangle =0$.
(The expression (2.10) is not complete form of superpotential
for the related flavons. 
We need some additional flavons for theoretical consistency.  
The form (2.10) is nothing but an outline expression. 
For the full expression of (2.10), see Ref.\cite{K-Y_PLB12}.)  
Since they assume that $\langle \Psi \rangle \gg \langle \Phi \rangle$,
they can neglect contributions to FGB masses from 
$\langle \Phi \rangle$, so that 
they obtain the inverse mass hierarchy of $M_{ij}$ (2.7). 

However, we have to note that only the loop diagram
$e_{i} \rightarrow e_{i}+A_i^{\ i} \rightarrow e_{i}$ 
can contribute to $m_{ei}(\mu)$ 
in Model A, while, in Model B, another loop diagrams 
$e_{i} \rightarrow e_j +A_i^{\ j} \rightarrow e_{i}$ 
with $j\neq i$ too can contribute to $m_{ei}(\mu)$
 in addition to 
$e_{i} \rightarrow e_i +A_i^{\ i} \rightarrow e_{i}$. 
This means that the original cancellation scenario 
$\delta_i \equiv \log m_{ei}^2 - \xi \log M_{ii}^2 = const$ 
is exchanged by 
$$
\delta_i \equiv \log m_{ei}^2 + \xi \sum_{j=1,2,3} \log M_{ij}^2
=  \log m_{ei}^2 + \xi \log S_i,
\eqno(2.11)
$$
where
$$
S_i \equiv M^2_{i1} M^2_{i2} M^2_{i3}
= K_e^3 \left(\frac{1}{m_{ei}} +\frac{1}{ m_{e1}} \right)  
\left(\frac{1}{m_{ei}} +\frac{1}{ m_{e2}} \right)
\left(\frac{1}{m_{ei}} +\frac{1}{ m_{e3}} \right) ,
\eqno(2.12)
$$
and $\xi$ is a family-number independent factor. 
From (2.11), the ratios $S_1/S_2$ and $S_2/S_3$ are given by
$S_1/S_2 \simeq (m_\mu/m_e)^2$ and 
$S_2/S_3 \simeq m_\tau/m_\mu$, so that the Sumino's 
cancellation mechanism in Model B does not hold exactly. 
Of course, the cancellation mechanism can be practically  
satisfied by adjusting a fine tuning parameter 
\cite{K-Y_PLB12}.
However, the cancellation mechanism with such a fine tuning 
parameter is not so beautiful compared 
with the original one (Model A).

\vspace{3mm}

\noindent{\bf 3  \  Basic idea}: 

In a new model (Model C), we change the VEV relation
(2.9) in Model B into
$$
\langle \Psi_i^{\ \alpha}\rangle 
\langle \bar{\Psi}_\alpha^{\ k}\rangle 
\left( \langle \Phi_k^{\ \beta}\rangle 
\langle \bar{\Phi}_\beta^{\ j}\rangle 
+ \delta_k^{\ j} \langle (\Phi_0)\rangle 
\langle (\bar{\Phi}_0)\rangle 
\right)
= k_\Psi k_\Phi \delta_i^j , 
\eqno(3.1)
$$
where we have assumed $R$ charges $R(\Phi_0) = R(\Phi)$. 
Here. we have introduced a new flavon $(\Phi_0)$ 
with $({\bf 1}, {\bf 1})$ of U(3)$\times$U(3)$'$
whose VEV is given by
$$
\langle \Phi_0 \rangle = v_0  .
\eqno(3.2)
$$
Then, we assume the following superpotential 
$$
W = \left[ (\Theta)_i^{\ j} (\Psi \bar{\Psi})_j^{\ k} 
(\Phi \bar{\Phi})_k^{\ i} \right] + 
\left( \left[ ((\Theta)_i^{\ j} (\Psi\bar{\Psi})_j^{\ i} \right]
-\left[ (\Theta)_i^{\ i} \right] \left[ (\Psi \bar{\Psi})_j^{\ j}
\right] \right) (\Phi_0 \bar{\Phi}_0) ,
\eqno(3.3)
$$
which leads to 
$$
\frac{(v_{F0})^2}{(v_{Fi})^2} = \frac{(v_{ei})^2 + (v_0)^2}{(v_0)^2}, 
\eqno(3.4)
$$
where $v_{F0}$ is defined by
$$
(v_{F0})^2 \equiv (v_{F1})^2+ (v_{F2})^2 +  (v_{F3})^2 . 
\eqno(3.5)
$$
However, the form (3.3) is not a general form.  
The form (3.3) is nothing but an ad hoc assumption. 
 
Hereafter, we denote the relation (3.4) as
$$
\frac{(v_{F0})^2}{(v_{Fi})^2} = \frac{m_{ei} + m_{0}} { m_{0}}, 
\eqno(3.6)
$$
where we put $m_{ei} = k_e (v_{ei})^2$ and 
$m_0 = k_e (v_0)^2$. 
Since we can express the left-hand side in Eq.(3.6) as
$$
\frac{(v_{F0})^2}{(v_{Fi})^2} 
= \frac{(v_{Fi})^2 +(v_{Fj})^2+ (v_{Fk})^2}{(v_{Fi})^2}
=  1 + \frac{(v_{Fj})^2 + (v_{Fk})^2}{(v_{Fi})^2} ,
\eqno(3.7)
$$
where  $(i,j,k)$ denotes cyclic permutation of $(1, 2, 3)$, 
we obtain a relation
$$
\frac{m_{ei}}{m_{0}}  = \frac{(v_{Fj})^2 + (v_{Fk})^2}{(v_{Fi})^2} .
\eqno(3.8)
$$

On the other hand, the factor $S_i$ is given by
$$
S_i = M_{i1}^2 M_{i2}^2 M_{i3}^2 = 2 (g_F^2)^3 (v_{Fi})^2 
[(v_{Fi})^2 +(v_{Fj})^2] [(v_{Fi})^2+ (v_{Fk})^2] = {A} 
\frac{(v_{Fi})^2}{(v_{Fj})^2 + (v_{Fk})^2} ,
\eqno(3.9)
$$
where $A \equiv 2 (g_F^2)^3 [(v_{F1})^2+(v_{F2})^2][(v_{F2})^2+(v_{F3})^2]
[(v_3^F)^2+(v_1^F)^2]$
is a family-number independent constant.
Thus, from Eq(3.8), we can express $S_i$ as
$$
S_i = A\, \frac{m_{0}}{m_{ei}} .
\eqno(3.10)
$$
Therefore, we get $\log S_i = - \log m_{ei} +\log (A\, m_{0})$, 
and we can achieve a complete Sumino cancellation mechanism. 

\vspace{3mm}

{\bf 4 \ Possible effects in the new scenario} 

Note that the value $m_0$ in Eq.(3.6) is not free under the 
observed values $(m_{e1}, m_{e2}, m_{e3})$. 
By using the relation (3.6), we obtain the following equation for $m_0$ 
$$
1=\frac{1}{(v_{F0})^2} [(v_{F1})^2+(v_{F2})^2)+(v_{F3})^2] =
\frac{m_{0}}{m_{e1} + m_{0}} + \frac{m_{0}}{m_{e2} + m_{0}} +
 \frac{ m_{0}}{m_{e3} + m_{0} }  .
\eqno(4.1)
$$
Then, Eq.(4.1) has only a positive solution $m_0 = 7.6219$ MeV, 
so that we can regard as
$m_{e1} \ll  m_{0} \ll m_{e2} \ll m_{e3}$. 
We can approximate Eq.(4.1) as
$$
\frac{1}{1+(m_{e1}/m_{0})} + \frac{m_{0}}{m_{e2}} +
 \frac{ m_{0}}{m_{e3}} \simeq 1 ,
\eqno(4.2)
$$
i.e.
$$
\frac{m_{0}}{m_{e2}} + \frac{ m_{0}}{m_{e3}}
\simeq \frac{m_{e1}}{m_{0}} \ \ 
\Rightarrow  \ \ 
(m_{0})^2 \simeq m_{e1} m_{e2} .
\eqno(4.3)
$$
Therefore, from (3.6) and (4.3), we obtain approximate relations
$$
(v_{F1})^2 \simeq (v_{F0})^2, \ \ \ (v_{F2})^2 \simeq \left(
\frac{\sqrt{m_{e1} m_{e2}} }{m_{e2}}\right) (v_{F0})^2, \ \ \ 
(v_{F3})^2 \simeq \left(\frac{\sqrt{m_{e1} m_{e2}} }{m_{e3}}
\right)^2 (v_{F0})^2. 
\eqno(4.4)
$$

Effect due to $m_0 \neq 0$ appears in values of 
FGB masses $M_{ij}$: 
As well as in Model B, the relative mass ratios of the FGBs 
are given \cite{YK_PLB14} by
$$
M_{33} : M_{32} : M_{22}: M_{31}: M_{21} : M_{11} =
1;  \sqrt{\frac{a^2+1}{2} }  : a: \sqrt{\frac{b^2+1}{2} } : 
\sqrt{ \frac{b^2+a^2}{2} } : b . 
 \eqno(4.5)
$$
However, differently form the case of Model B, where 
$a \equiv M_{22}/M_{33}$ and $b\equiv M_{11}/M_{33}$, 
the parameters $a$ and $b$ in Model C are given by
 $$
 a\equiv \frac{M_{22}}{M_{33.} }= \frac{v_{F2}}{v_{F3}} =
\left( \frac{m_{e3} + m_{0}}{m_{e2}+ m_{0}} \right)^{1/2} , \ \ \ \ 
 b\equiv \frac{M_{11}}{M_{33} } = \frac{v_{F1}}{v_{F3}} =
\left( \frac{m_{e3} + m_{0}}{m_{e1}+m_{0}} \right)^{1/2} . 
\eqno(4.6)
$$
(The values in Model B are obtained by putting  $m_0 =0$ in Eq.(4.6). )
Let us show numerical values $a$ and $b$ in the present model (Model C) 
without approximation (4.3): 
$$
\begin{array}{lc}
 a^C=3.97347, & b^C=15.0691, \\
 (a^B =4.10081, & b^B=58.9674), \\
\end{array}
\eqno(4.7)  
$$
 Here, for the sake of comparison, we have also shown values 
 $a$ and $b$ in Model B as $a^B$ and $b^B$, respectively. 
The value $a^C$ is almost same as the value $a^B$, while 
 the value $b^C$ is considerably smaller than the value $b^B$,
$b^C \sim \frac{1}{4} b^B$. 

Lower bounds of $M_{ij}$ are constrained 
by the observed $P^0$-$\bar{P}^0$ mixing ($P=K, D, B, B_s$).
The numerical results in Model B have been given   
in Ref.\cite{YK_PLB14}.  
(However, take notice that Model B in the present paper 
corresponds to ``Model A$_1$" in Ref.\cite{YK_PLB14}.) 
As seen in Eq.(4.5), the values $M_{33}$, $M_{32}$ and
$M_{22}$ are independent of the parameter $b$, so that 
those values will almost be unchanged under the change of 
the value $m_0$.   
On the other hand, in a model with an inverse FGB masses, 
the constraint from the $P^0$-$\bar{P}^0$ mixing is almost 
determined by the value $M_{22}$ (and $M_{33}$ for 
$B_s^0$-$\bar{B}_s^0$ mixing) \cite{YK_PLB14}, 
so that the values of $M_{22}$, $M_{23}$ and $M_{33}$ are 
almost unchanged in Model C. 
(Exactly speaking, the constraint is fixed by a form of 
effective mass $\tilde{M}_{ij} \equiv M_{ij}/(g_F/\sqrt2)$.)
Therefore, we approximately put
$$
\tilde{M}_{22}^C \equiv \frac{M_{22}^C}{g_F^C/\sqrt2} 
\simeq \tilde{M}_{22}^B \equiv \frac{M_{22}^B}{g_F^B/\sqrt2} .
\eqno(4.8)
$$
In Model B \cite{K-Y_PLB12}, 
a cancellation condition $(g_F/\sqrt2)^2=(3/2) \zeta e^2$
has taken, where $\zeta$ is a fine tuning factor because
the Sumino's cancellation works only approximately, and 
the value $\zeta=1.752$ was taken. 
On the other hand, in Model C, the Sumino's cancellation 
holds exactly, so that we have to put $\zeta=1$. 
Therefore, the value $g_F^C/g_F^B$ is given by
$$
\frac{g_F^C}{g_F^B} = \frac{1}{\sqrt{\zeta}} = 
\frac{1}{\sqrt{1.752}} .
\eqno(4.9)
$$ 
Then. we can estimate the values of $M_{ij}^C$ in Model C
from the numerical values provided in Ref.\cite{YK_PLB14}. 
The results are given Table 1.

For the sake of comparison, correspondingly to Ref.\cite{YK_PLB14}, 
we have added a case of $n=2$ in addition to a case $n=1$, 
where $n$ is defined by 
$$
M_{ij}^2 = K_e (m_{ei}^{-n} + m_{ej}^{-n}) .
\eqno(4.10)
$$
The extension to $n=2$ is also possible, although 
it needs somewhat complicated framework compared with 
Model B.
For Case $n=2$, we obtain parameter values 
$a^C = 16.7763$ and $b^C= 241.447$, 
($a^B= 16.8167$ and $b^B =3477.15$), 
correspondingly to Eq.(4.7).  
Also, we use $(g_F^{n=1}/\sqrt{2})^B=0.4339$ and 
$(g_F^{n=2}/\sqrt{2})^B=0.3068$. 

\vspace{2mm}
\begin{center}
Table 1 \ Lower bound of $M_{ij}$ [TeV] from the observed 
$P^0$-$\bar{P}^0$ mixing 

\vspace{2mm}
\begin{tabular}{|cc|cccccc|} \hline
Model & $n$  & $M_{11}$ &  $M_{12}$ &  $M_{13}$ & 
 $M_{22}$ &  $M_{23}$ &  $M_{33}$ \\[0.05in]\hline
Model C & $n=1$ & 4.38$\times 10^2$ &  3.20$\times 10^2$ & 
3.10$\times 10^2$ & 115 & 84.1 & 29.0 \\
Model B & $n=1$ &  2.20$\times 10^3$ &  1.56$\times 10^3$ & 
1.55$\times 10^3$ & 153 &  111 & 37.2 \\[0.05in]\hline        
Model C & $n=2$ &  1.18$\times 10^3$  & 0.839$\times 10^3$  & 
0.837$\times 10^3$  & 82.1 & 58.2 & 4.90 \\
Model B & $n=2$  &  2.25$\times 10^4$ & 1.59$\times 10^4$ & 
1.59$\times 10^4$ & 109 & 77.1 & 6.47 \\[0.05in]\hline
\end{tabular}
\end{center}
\vspace{2mm}
       
As seen in Table 1, we can obtain somewhat lower mass 
values compared with the previous values (in Model B) 
only for $M_{11}$, $M_{12}$ and $M_{13}$.
However, the FGBs $A_1^{\ 1}$, $A_2^{\ 1}$ and $A_3^{\ 1}$ 
have masses of $10^{2-3}$ TeV scale, it is not easy 
to observe the effects due to $m_0 \neq 0$. 
For example,  let us see expect rare 
decays $K^+ \rightarrow \pi^+ \mu^+ e^-$ via $A_1^{\ 2}$ 
and $B^+ \rightarrow \pi^+ \tau^+ e^-$ via $A_1^{\ 3}$, 
which are proportional to $(\tilde{M}_{ij})^{-4} \equiv 
(g_F/M_{ij})^4$.  
From Table 1 and Eq.(4.9), we obtain
$$
\left( \frac{\tilde{M}_{12}^B}{\tilde{M}_{12}^C} \right)^4 
\simeq 1.84 \times 10^2, \ \ \ 
\left( \frac{\tilde{M}_{13}^B}{\tilde{M}_{13}^C} \right)^4 
\simeq 2.04 \times 10^2, \ \ \ \ (n=1) ,
\eqno(4.11)
$$
for $n=1$.
The values (4.11)are still  
insufficient to observe such rare decays. 
However, if we consider such observations for the
case $n=2$, we obtain 
$$
\left( \frac{\tilde{M}_{12}^B}{\tilde{M}_{12}^C} \right)^4 
\simeq 4.20 \times 10^4, \ \ \ 
\left( \frac{\tilde{M}_{13}^B}{\tilde{M}_{13}^C} \right)^4 
\simeq 4.24 \times 10^4, \ \ \ \ (n=2), 
\eqno(4.12)
$$
The predicted value in Model B was 
$Br(K^+ \rightarrow \pi^+ e^- \mu^+) 
 \simeq 2.3\times 10^{-16}$ 
 \cite{K-Y_PLB12}. 
On the other hand, we can predict
$$
 Br(K^+ \rightarrow \pi^+ e^- \mu^+) \simeq 0.97 \times 10^{-11} ,
\eqno(4.13)
$$
in Model C.
Since the experiments have reported \cite{PDG16} 
$Br(K^+ \rightarrow \pi^+ e^- \mu^+) < 1.3 \times 10^{-11}$,
the observation is promising in the near future. 
We also expect observation of $B^+ \rightarrow \pi^+ \tau^+ e^-$
in the near future. 
(Of course the values $M_{ij}$ in Table 1 are lower bound of FGB masses 
from the observation of $P^0$-$\bar{P}^0$ mixing, so that 
it is not likely that the actual masses are coincidentally the same 
as the lower bounds from the $P^0$-$\bar{P}^0$ mixing.)  
We hope further investigation of rare decays.  
Besides, the case $n=2$ predicts the lightest FGB mass 
as $M_{33}=4.9$ TeV. 
This value is within reach of the 14 TeV LHC experiments, 
i.e. $p+p \rightarrow A_3^{\ 3} +X \rightarrow \tau^+ \tau^- +X$. 
 
The most promising visible effect will appear 
the so-called $\mu$-$e$ conversion experiments 
$$
R(N) \equiv \frac{\sigma(\mu^- N \rightarrow e^- N)}{
\sigma(\mu\  {\rm capture}) }.
\eqno(4.14)
$$
For example, the COMET experiment \cite{COMET} aims 
for a goal $R \sim 10^{-17}$. 
On the other hand, we roughly estimate $R({\rm Al})$ as
$$
R({\rm Al}) \simeq 0.85 \times 10^{-16} \frac{1}{n^2} 
\left( \frac{10^3 [{\rm TeV}] }{
{M}_{12} [{\rm TeV}] } \right)^4 ,
\eqno(4.15)
$$
in Models B and C.  
(For $\mu$-$e$ conversion induced by exchange of the FGB $A_2^{\ 1}$, 
for example, see Ref.\cite{mu-e-conv_FGB}.)
Since the value of $M_{12}^B$ in Model B \cite{mu-e-conv_FGB}
$n=1$ is $1.56 \times 10^3$ TeV, so that the value gives 
$R({\rm Al}) \sim 1.44 \times 10^{-17}$,  the observation 
of $\mu$-$e$ conversion was critical in Model B.
On the other hand, in Model C, the revised value (4.11) in the 
case $n=1$ can give $R({\rm Al})^C \sim 2.7 \times 10^{-15}$. 
Therefore, the observation of $\mu$-$e$ conversion due to FGB is 
promising in the experiments
\cite{COMET, Mu2e}, 
even if we take into consideration
that the mass values in Table 1 are nothing but lower bounds 
constrained from the observed $P^0$-$\bar{P}^0$ mixing. 
 
\vspace{3mm}

{\bf 5 \ Concluding remarks} 

In conclusion, we have proposed a revised FGB model of
Model B \cite{K-Y_PLB12} with an inverted FGB mass hierarchy. 
The model (Model C) is anomaly free as well as Model B, 
while the Sumino cancellation mechanism exactly holds 
as well as the original Sumino model (Model A).\footnote{
As well as the Sumino model, this does not mean that 
the relation $K(\mu) =2/3$ holds in whole energy scale range. 
As we noted already, there are many neglected contributions 
to $K(\mu)$ in the Sumino model, so that Sumino has speculated 
that the relation $K(\mu) = 2/3$ holds only within an energy 
scale range smaller than $10^3$ TeV. 
Our result (the largest FGB mass $M_{11} \sim 10^3$ TeV) 
in Table 1 is consistent with the Sumino's speculation. 
}
Therefore, we have obtained a FGB model, where 
all merits in the original Sumino model are retained, but 
all problems in the original model disappear.

The purpose of the present paper is to improve the formulation 
of the Model B. 
(The improvement of Model A has been already done by Model B.)    
As a result, we have obtained considerably 
low mass scales of the FGBs $A_2^{\ 1}$ and $A_3^{\ 1}$ 
compared with Model B. 
If we suppose that the nature chooses not always a simple case,
the case $n=2$ is very interesting from the phenomenological 
point of view, for example, $K^+ \rightarrow \pi^+ \mu^+ e^-$ 
and so on. 
(However, we need somewhat complicated model-building for 
the case $n=2$.)  
On the other hand, 
if we suppose that the nature choose the simplest case $n=1$,  
the visible effect will be an observation in the 
$\mu^- N \rightarrow e^- N$ experiments.  
Those observations are within our reach.  

Present results on $M_{ij}$ rely on the validity of 
the Sumino cancellation mechanism.
If we leave the Sumino FGB model, we can take 
any values of $M_{ij}$ (and also any value of $g_F$)  
from the theoretical point of view. 
However, then, the values $M_{ij}$ will be severely 
restricted from the observed $P^0$-$\bar{P}^0$ mixing.  
We again  would like to emphasize the following 
point: The Sumino cancellation mechanism has an applicable 
energy scale range. 
The validity of the Sumino mechanism is confirmed not 
by seeing how the formula (1.1) is excellently satisfied 
with the pole masses, but by seeing that, at what 
energy scale, the well-satisfied relation $K^{pole}=2/3$ 
breaks down.

We are convinced that the basic idea for a realistic FGB model
by Sumino should be taken seriously. 
We hope that FGBs become more realistic and more
familiar to us.


\vspace{5mm}

\noindent{\large\bf  \ Acknowledgments} 

The work was supported by JSPS KAKENHI Grant
number JP16K05325. 
The author thanks Y.~Sumino and T.~Yamashita for helpful discussion 
on the effective superpotential and helpful comments. 


\vspace{5mm}

%

%

\begin{thebibliography}{99} 
%
\bibitem{Sumino_PLB09}
  Y.~Sumino,
  Phys.\ Lett.\  B {\bf 671}, 477 (2009). 
%
\bibitem{K-mass} 
  Y.~Koide,
  Lett.\ Nuovo Cim.\  {\bf 34}, 201 (1982);
  Phys.\ Lett.\  B {\bf 120}, 161 (1983);
%
  Phys.\ Rev.\  D {\bf 28}, 252 (1983).
%
\bibitem{K-Y_PLB12}
Y.~Koide and T.~Yamashita, 
  Phys.\ Lett.\  B {\bf 711}, 384 (2012).
%
%
\bibitem{YK_MPLA90}
Y.~Koide, Mod.~Phys.~Lett. A {\bf 28}, 2319 (1990). 
%
%
\bibitem{Ma_PLB07} E.~Ma, Phys. Lett. B {\bf 649}, 287 (2007).
%
\bibitem{Sumino_JHEP09} Y.~Sumino, JHEP {\bf 0905}, 075 (2009).
%
\bibitem{Arason92}
H.~Arason, {\it et al}., Phys.~Rev. {\bf D 46}, 3945 (1992).
\bibitem{Sumino_private} Y.~Sumino, private communication.
%
\bibitem{YK_PLB14} 
Y.~Koide,  Phys.\ Lett.\  B {\bf 736}, 499 (2014).
%
\bibitem{PDG16}
  C.~Patrignani {\it et al.}  [Particle Data Group Collaboration],
  Chin.\ Phys.\ C {\bf 40} (2016) 100001.
 %
%
\bibitem{COMET} 
Y.~Kuno, Prog.~Theor.~Exp.~Phys. 022C01 (2013);
  Y.~G.~Cui {\it et al.}  [COMET Collaboration],
  KEK-2009-10.
%
%
\bibitem{mu-e-conv_FGB} Y.~Koide and M.~Yamanaka, 
  Phys.\ Lett.\  B {\bf 762}, 41 (2016).
%
\bibitem{Mu2e}
L.~Bartoszek {\it et al.} [Mu2e Collaboration],
  arXiv:1501.05241 [physics.ins-det].
%
\end{thebibliography}
\end{document}